# HYBRID SCENARIO BASED PERFORMANCE ANALYSIS OF DSDV AND DSR

Koushik Majumder [1] and Subir Kumar Sarkar[2]

[1] Department of Computer Science and Engineering, West Bengal University of Technology, Kolkata, INDIA
[2] Department of Electronics and Telecommunication Engineering, Jadavpur University, Kolkata, INDIA
koushikmail@yahoo.com

## ABSTRACT

*The area of mobile ad hoc networking has received considerable attention of the research community in recent years. These networks have gained immense popularity primarily due to their infrastructure-less mode of operation which makes them a suitable candidate for deployment in emergency scenarios like relief operation, battlefield etc., where either the pre-existing infrastructure is totally damaged or it is not possible to establish a new infrastructure quickly. However, MANETs are constrained due to the limited transmission range of the mobile nodes which reduces the total coverage area. Sometimes the infrastructure-less ad hoc network may be combined with a fixed network to form a hybrid network which can cover a wider area with the advantage of having less fixed infrastructure. In such a combined network, for transferring data, we need base stations which act as gateways between the wired and wireless domains. Due to the hybrid nature of these networks, routing is considered a challenging task. Several routing protocols have been proposed and tested under various traffic conditions. However, the simulations of such routing protocols usually do not consider the hybrid network scenario. In this work we have carried out a systematic performance study of the two prominent routing protocols: Destination Sequenced Distance Vector Routing (DSDV) and Dynamic Source Routing (DSR) protocols in the hybrid networking environment. We have analyzed the performance differentials on the basis of three metrics – packet delivery fraction, average end-to-end delay and normalized routing load under varying pause time with different number of sources using NS2 based simulation.*

## 1. INTRODUCTION

In recent years there has been a huge influx of laptops, handheld computers, PDAs and mobile phones in our daily lives and the industry has seen tremendous growth in the wireless arena. With recent advances in technology these devices are becoming increasingly popular as a result of their decreasing cost, higher processing capability, greater storage capacity, smaller size and their support for newer and wider range of applications. Due to their small sizes and battery powered operation, users can move with these devices freely without being restricted to one place. To combat this huge flow highly portable devices the mobile ad hoc networks (MANET) [1-14] came into being. These networks are ad hoc because there is no fixed infrastructure or centralized server support.

Each node acts both as the host as well as the router. The nodes which are within the communication range of each other can directly communicate between them. But, if a source node wants to send data to a destination node, which is outside of its communication range, in that case it has to forward the data packet through intermediate nodes. Communication in ad hoc network is thus multi-hop.

In the early 1970s DARPA sponsored the earliest wireless ad hoc networks called "packet radio" networks (PRNET) [15]. The Ham radio community also made similar experiments in this field. These earlier versions of the packet radio networks predated the Internet and indeed were the part of the motivation of the original Internet Protocol suite. Later in the 1980s DARPA made experiments in the Survival Radio Network (SURAN) [16] project. The third wave of academic activities began in the mid 1990s with the advent of the inexpensive 802.11 radio cards for personal computers.

The mobile ad hoc networks are becoming increasingly popular due their less costly and rapid deployability, inherent support for mobility and the potential to provide ad hoc connectivity to devices thus allowing them to form temporary networks in areas where either there is no fixed infrastructure or the pre-existing infrastructure is totally destroyed and it is not possible to establish a new infrastructure quickly. Due to these features mobile ad hoc networks are receiving increasing attention in areas such as – real time information sharing between soldiers in the battle field, emergency natural disaster relief operation, students participating in an interactive lecture session with their hand-held computers etc.

Routing in mobile ad hoc network is considered a challenging task due to the drastic and unpredictable changes in the network topology resulting from the random and frequent movement of the nodes and due to the absence of any centralized control. Several routing protocols have been developed under the aegis of Mobile Ad hoc Networking (MANET) working group, which is a charter of Internet Engineering Task Force (IETF).

The routing protocols for ad hoc networks can be divided into two broad categories: proactive and reactive. In protocols following the proactive approach like DSDV [17], CGSR [18], STAR [19], OLSR [20], HSR [21], GSR [22] it is necessary for the nodes in the ad hoc network to maintain consistent routing information from each node to all other nodes. In order to keep the information up-to-date, the nodes need to exchange the routing information periodically. The main advantage of this type of protocols is - whenever a node wants to send a data packet to another destination node, it can do that without wasting any time for path setup.

In case of reactive routing protocols such as DSR [23, 24], AODV [25, 26], ABR [27], SSA [28], FORP [29], PLBR [30] a lazy approach is followed. Here the nodes need not maintain the routes to all other nodes. Routes to the destinations are determined by flooding the whole network with route query packets only when required. The main advantage of this type of protocols is – a lot of precious network bandwidth can be saved as periodic route exchanges are no longer needed.

In order to claim the advantage from both of these types, protocols like CEDAR [31], ZRP [32], and ZHLS [33] combine both the proactive and the reactive approach.

Sometimes a hybrid network can be formed by combining the ad hoc network with the wired network. In these hybrid networks data can move from the mobile to the non-mobile nodes and vice-versa. We need the base stations for this purpose, which act as the gateways between the wired and wireless domains. By using this combination we can cover a larger area with less fixed infrastructure, less number of fixed antennas and base station and can reduce the overall power consumption.

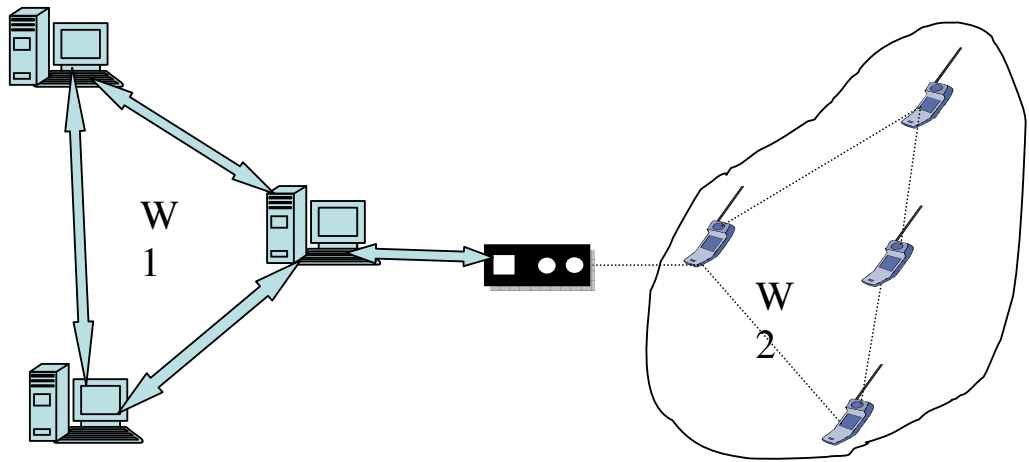

Figure 1. Hybrid network of wired domain W1 and wireless domain W2

Several simulation based experiments have been made to compare the performance of the routing protocols for mobile ad hoc network. Das et al.[34] made performance comparison of routing protocols for MANET based on the number of conversations per mobile node. They used the Maryland Routing Simulator(MARS) for this purpose. Performance comparison results of two on demand routing protocols – AODV and DSR is presented in the work of Das, Perkins and Royer [35]. They used NS2 based simulation. CBR sources were used with packet size of 512 bytes. Two different simulation set ups were used. One with 50 nodes and 1500m x 300m simulation area and the other with 100 nods and 2200m x 600m simulation area. The performance metrics studied were : packet delivery fraction, average end-to-end delay and normalized routing load. Broch, Maltz, Johnson, Hu and Jetcheva have investigated the performance of four different ad hoc routing protocols AODV, DSDV, DSR and TORA in their work [36] using NS2 based simulation. The simulations were carried out over a rectangular area of 1500m x 3000m with 50 nodes and for a period of 900 seconds. The random waypoint model was selected as the movement model. Packet delivery ratio, routing overhead and path optimality – these metrics were considered for performance evaluation. These works, however, do not take into consideration the influence of hybrid network scenario over the performance of the routing protocols. On the contrary, in this paper we carry out a systematic performance evaluation of the two routing protocols for mobile ad hoc network – Destination Sequenced Distance Vector Routing (DSDV) and Dynamic Source Routing (DSR) protocol in the hybrid networking environment. We have used the means of simulation using NS-2[37, 38, 39,40] to gather data about these routing protocols in order to evaluate their performance.

The rest of the paper is organized as follows. Section 2 gives a brief introduction of Dynamic Source Routing (DSR) and Destination Sequenced Distance Vector Routing (DSDV) protocols. We describe the simulation environment for the hybrid scenario in Section 3. Section 4 details the key performance metrics used in the study. In Section 5 we present the simulation results and analysis of our observation. Finally Section 6 concludes the paper and defines topics for further research.

## 2. DESCRIPTION OF ROUTING PROTOCOLS

### 2.1. Dynamic Source Routing (DSR)

The Dynamic Source Routing Protocol (DSR) [23,24] is a reactive routing protocol. The main feature of DSR is the use of source routing technique. In this technique the source node knows the complete hop-by-hop route towards the destination node. The source node lists this entire sequence in the packet's header. If a node wants to send a packet to a destination, the route to which is unknown, in that case a dynamic route discovery process is initiated to discover the route. DSR consists of the Route Discovery and Route Maintenance phase, through which it discovers and maintains source routes to arbitrary destinations in the network.

### 2.1.1. Route Discovery

If a node A wants to send a packet to a destination node B, it searches its Route Cache. If the Route Cache contains a valid route, node A inserts this route into the header of the packet and sends the data packet to the destination B. In case when no route is found in the Route Cache, a Route Discovery is initiated.

Node A initiates the Route Discovery by broadcasting a ROUTE REQUEST message. All nodes within the transmission range receive this message. The nodes which are not in the route, add their address to the route record in the packet and forward the packet when received for the first time. They check the request id and source node id to avoid multiple retransmissions. The destination node B sends a ROUTE REPLY when it receives a ROUTE REQUEST. If the link is bidirectional, the ROUTE REPLY propagates through the reverse route of the ROUTE REQUEST. If the link is unidirectional, in that case B checks its own Route Cache for a route to A and uses it to send the ROUTE REPLY to the source A. If no route is found, B will start its own Route Discovery. In order to avoid infinite numbers of Route Discoveries it piggybacks the original ROUTE REQUEST message to its own. The route information carried back by the ROUTE REPLY message is cached at the source for future use. In addition to the destination node, other intermediate nodes can also send replies to a ROUTE REQUEST using cached routes to the destination.

### 2.1.2. Route Maintenance

The node which sends a packet using a source route is responsible for acknowledging the receipt of the packet by the next node. A packet is retransmitted until a receipt is received or the maximum number of retransmissions is exceeded. If no confirmation is received, the node transmits a ROUTE ERROR message to the original sender indicating a broken link. The ROUTE ERROR packet causes the intermediate nodes to remove the routes containing the broken link from their route caches. Ultimately the sender will remove this link from its cache and look for another source route to the destination in its cache. If the route cache contains another source route, the node sends the packet using this route. Otherwise, it needs to initialize a new route discovery process. DSR makes very effective use of source routing and route caching. In order to improve performance any forwarding node caches the source route contained in a packet forwarded by it for possible future use.

### 2.2. Destination Sequenced Distance Vector Routing (DSDV)

The Destination Sequenced Distance Vector Routing (DSDV) [17] is a proactive or table driven routing protocol designed for MANET. It was developed by C. Perkins and P. Bhagwat in 19994. This scheme is based on the classical Bellman-Ford distance vector algorithm with certain modifications to make it suitable for the ad hoc environment and to solve the problem of routing loop and count-to-infinity. In DSDV every node maintains a routing table which contains the list of all possible destinations within the network and the number of hops to reach each possible destination. Each distance entry is marked by a sequence number usually originated by the destination node. This sequence numbering scheme is used to counter the count-to infinity problem and to distinguish the stale routes from the fresh ones thus avoiding the formation of loops.

In order to maintain up-to-date routing information about the frequently changing topology of the network the nodes need periodic exchanges of routing tables with their neighbours. But this will create a huge overhead of control packets in an already bandwidth constrained network. To reduce this huge overhead of control traffic the routing updates are generally classified into two types – full dump and incremental update. In case of full dumps, nodes need to exchange complete routing tables with their neighbours. Full dumps are needed to maintain consistent routing information when the network topology changes completely and very fast due to frequent movement of nodes. But this may result in a large number of routing packet exchanges between the nodes. On the other hand incremental updates contain only those entries that have been updated since the last full dumps. Incremental updates are much smaller in size than the full dumps and they should fit in a single Network Protocol Data Unit (NPDU). When the network is relatively stable, incremental updates are used to rapidly propagate the routing information regarding the small changes in network topology. This saves a lot of network traffic. In addition to the periodic updates DSDV uses triggered updates, when significant new information is available about the topological change. Thus the update is both time-driven as well as event-driven.

Table updates are initiated by the destination nodes and they generate the sequence numbers. Every node periodically transmits their routing updates to their immediate neighbours with monotonically increasing sequence numbers. After receiving a new route update, every node compares it with its existing entry. Routes with smaller sequence numbers are simply discarded and the one with the recent sequence number is selected. In case when the new route is having the same sequence number as the existing route, the one with the smaller hop count is selected. If the new route is chosen, its hop-count is incremented by one, as the packets will require one more hop to reach the destination. This change in the routing information is then immediately communicated to the neighbours.

When a node S finds that its route to destination D is broken, it advertises its link to destination D with an infinite hop-count and a sequence number that is one greater than the sequence number of the broken route. This is the only case when the sequence number is not assigned by the destination node. Sequence numbers defined by the originating nodes are even numbers, whereas the sequence numbers indicating the broken links are odd numbers. After having this infinite hop-count entry, when a node, later receives a finite hop-count entry with newer sequence number, it immediately broadcasts its new routing update. The broken links are thus quickly replaced by the real routes.

## 3. SIMULATION MODEL

We have done our simulation based on ns-2.34 which has the support for the simulation of multi-hop wireless ad hoc network completed with physical, data link and medium access control(MAC) layer models. NS is a discrete event simulator. It was developed by the University of California at Berkeley and the VINT project [37]. Our main goal was to measure the performance of the protocols under a range of varying network conditions. We have used the Distributed Coordination Function (DCF) of IEEE 802.11[41] for wireless LANs as the MAC layer protocol. DCF uses RTS/CTS frame along with random backoff mechanism to resolve the medium contention conflict. Data packets were transmitted using an unslotted carrier sense multiple access (CSMA) technique with collision avoidance (CSMA/CA)[41]. We have used the radio model whose characteristics are similar to the Lucent WaveLAN [42] direct sequence spread spectrum radio.

As buffering is needed for the data packets which are destined for a particular target node and for which the route discovery process is currently going on, the protocols have a send buffer of 64 packets. In order to prevent indefinite waiting for these data packets, the packets are dropped from the buffers when the waiting time exceeds 30 seconds. The interface queue has the capacity to hold 50 packets and it is maintained as a priority queue. The interface queue holds both the data and control traffic sent by the routing layer until they are transmitted by the MAC layer. The control packets get higher priority than the data packets.

### 3.1. Mobility Model

Inclusion of a mobility model is necessary in order to evaluate the performance of a protocol for ad hoc network in a simulated environment. Here in our work we have used the random waypoint [43] model. This model is a simple and common mobility model and is widely used for the performance evaluation of MANET protocols in simulated environment. This particular mobility model has pause time between changes in direction and/or speed. The mobile nodes are initially distributed over the entire simulation area. In order to ensure randomness in the initial distribution data gathering has to start after a certain simulation time. A mobile node starts simulation by waiting at one location for a specified pause time. After this time is over, it randomly selects the next destination in the simulation area. It also chooses a random speed uniformly distributed between a maximum and minimum speed and travels with a speed $v$ whose value is uniformly chosen in the interval $(0, v_{max})$. Then the mobile node moves towards its selected destination at the selected speed. After reaching its destination, the mobile node again waits for the specified pause time before choosing a new way point and speed.

### 3.2. Movement Model

In the simulation environment the nodes move according to our selected random waypoint mobility model. We have generated the movement scenario files using the *setdest* program which comes with the NS-2 distribution. These scenario files are characterized by pause time. The total duration of our each simulation run is 800 seconds. We have varied our simulation with movement patterns for nine different pause times: 0, 100, 200, 300, 400, 500, 600, 700, 800 seconds. These varying pause times affect the relative speed of the mobile nodes. A pause time of 800 seconds corresponds to the motionless state of the nodes in the simulation environment as the total duration of the simulation run is 800 seconds. On the contrary when we choose the pause time of 0 second it indicates continuous motion of the nodes. We have performed our experiment with two different numbers of source nodes: 15 source nodes and 25 source nodes. As slight changes in the movement pattern will have significant effect on the protocol performance, we have generated scenario files with 90 different movement patterns, 10 for each value of pause time. In order to compare the performance of the protocols based on the identical scenario both the protocols were run with these 90 different movement patterns.

### 3.3. Communication Model

In our simulation environment the MANET nodes use constant bit rate (CBR) traffic sources when they send data to the wired domain. We have used the cbrgen traffic-scenario generator tool available in NS2 to generate the CBR traffic connections between the nodes. Data packets transmitted are of 512 bytes. We have used two different communication patterns corresponding to 15 and 25 sources. Data packets are sent at the rate of 4 packets/second. All communications are peer-to-peer in these patterns.

### 3.4. Hybrid Scenario

We have used a rectangular simulation area of 800 m x 500 m. The choice of rectangular area instead of square area was made in order to ensure longer routes between nodes. In our simulation we have used two ray ground propagation model. Our mixed scenario consists of a wireless and a wired domain. The simulation was performed with 50 wireless nodes and 10 wired nodes. For our hybrid network environment we have a base station located at the centre (400,250) of the simulation area. The base station acts as a gateway between the wireless and wired domains. Every communication between the wired and wireless part goes through the base station. For our mixed simulation scenario we have turned on hierarchical routing in order to route packets between the wired and the wireless domains. The domains and clusters are defined by using the hierarchical topology structure. As the base station nodes act as gateways between the wired and wireless domain they need to have their wired routing on. In the simulation setup we have done this by setting the node-config option –wiredRouting on. After the configuration of the base station the wireless nodes are reconfigured by turning their wiredRouting off. If a wired node wants to send a packet to a wireless node, the packet is first sent to the base station. The base station then uses its ad hoc routing protocol to send the packet to its correct destination. Similarly the packets that originate from the mobile nodes and are destined outside the wireless domain are forwarded by the mobile nodes towards their assigned base station. Base station then forwards the packets towards the wired domain.

## 4. PERFORMANCE METRICS

We have primarily selected the following three performance metrics in order to study the performance comparison of DSDV and DSR.

**Packet delivery fraction**: This is defined as the ratio between the number of delivered packets and those generated by the constant bit rate (CBR) traffic sources.

**Average end-to-end delay**: This metric includes all possible delays caused by buffering at the time of the route discovery, queuing delay due to waiting at the interface queue, retransmission delays at MAC, propagation and transfer times. This is basically defined as the ratio between the summation of the time difference between the packet received time and the packet sent time and the summation of data packets received by all nodes.

**Normalized routing load**: This is defined as the number of routing packets transmitted per data packet delivered at the destination. Each hop-wise transmission of a routing packet is counted as one transmission.

## 5. SIMULATION RESULTS AND ANALYSIS

The following table summarizes the simulation parameters that we have selected in order to evaluate the performance of the two routing protocols.

Table 1. Simulation parameters.

| Simulation Parameters | |
| --- | --- |
| Protocols | DSDV, DSR |
| Number of mobile nodes | 50 |
| Number of fixed nodes | 10 |
| Simulation area size | 800 m x 500 m |
| Simulation duration | 800 seconds |
| Mobility model | Random way point |
| Traffic type | Constant bit rate(CBR) |
| Packet size | 512 bytes |
| Max speed | 20m/sec |
| Connection rate | 4packets/sec |
| Pause time | 0, 100, 200, 300, 400, 500, 600, 700, 800 |
| Number of sources | 15,25 |

## 5.1. Packet Delivery Fraction (PDF) Comparison

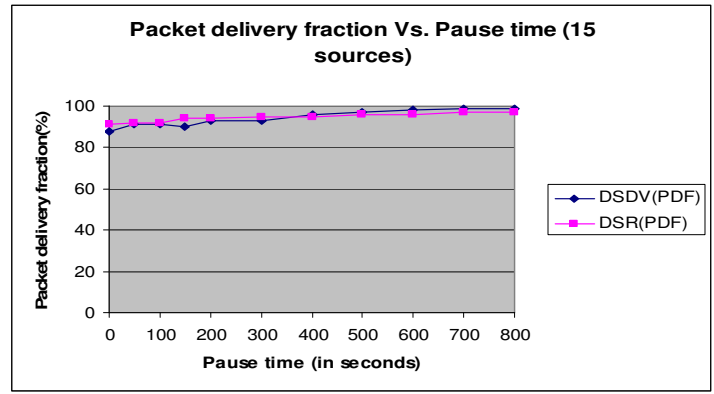

Figure 2. Packet Delivery Fraction vs. Pause Time for 15 sources

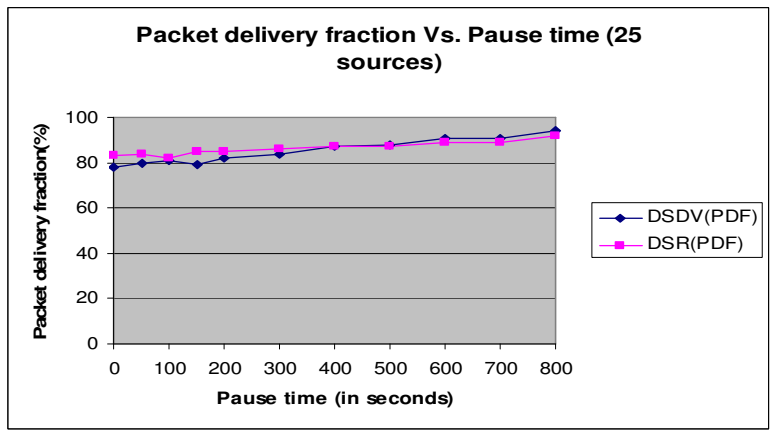

Figure 3. Packet Delivery Fraction vs. Pause Time for 25 sources

Figure2 and Figure3 show the packet delivery performances of DSDV and DSR from our simulation experiments. We have measured the packet delivery fraction of these two protocols by varying the pause time with respect to 15 and 25 numbers of sources. From the graphs we see that DSDV shows better packet delivery performance than DSR at lower mobility. This is due to the fact that, at low mobility all the routes are already available due to the proactive nature of DSDV. Therefore, most of the packets will be delivered smoothly. Whereas, DSR, being a source routing protocol, a significant time will be required for initial path setup. During this time, no packets can be delivered to the destination due to unavailability of routes.

With high mobility there will be frequent and high volume of changes in the network topology. The proactive nature of DSDV makes it less adaptive to this frequent change. In DSDV, with these major changes in network topology, greater number of full dumps needs to be exchanged between the nodes in order to maintain up-to-date routing information at the nodes. This huge volume of control traffic consumes a significant part of the channel bandwidth and lesser channel capacity is left for the data traffic which results in reduced packet delivery fraction of DSDV at higher mobility. Moreover, in DSDV packets are dropped due to stale routing table entry. DSDV keeps track of only one route per destination. Due to lack of alternate routes, MAC layer drops packets that it is unable to deliver through stale routes. DSR on the contrary, is more adaptive to the frequently changing scenario due to its on-demand routing nature. In case of DSR, multiple routes exist in the cache. Thus, even if a link is broken due to high mobility, alternative routes can be found from the cache. This prevents packet dropping and results in better packet delivery performance of DSR.

In both the graphs we see that as the mobility and number of sources increase, the packet delivery performance of both these protocols decreases. This happens due to the fact that with increasing mobility and greater number of sources, finding the route requires more and more routing traffic thus leaving a lesser portion of the channel available for network data traffic. Moreover, with reduced pause time as the network topology changes frequently, more number of routes becomes stale quickly. But the source node having no mechanism to determine a stale route, uses the same stale route to forward the packet. This causes more and more number of packets to be dropped.

### 5.2. Average End-to-End Delay Comparison

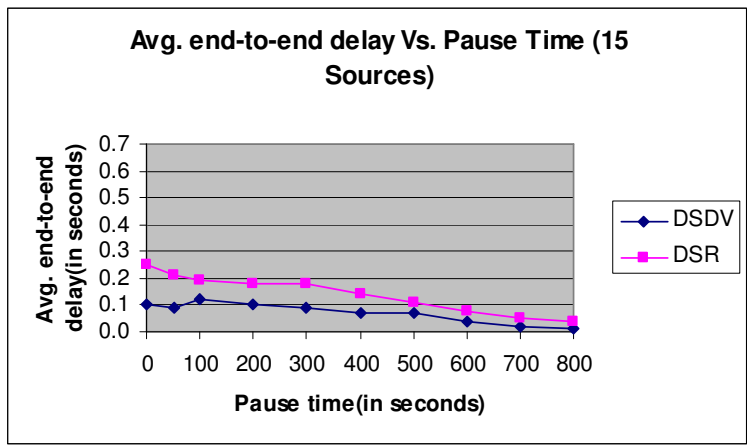

Figure 4. Average End to End Delay vs. Pause time for 15 Sources

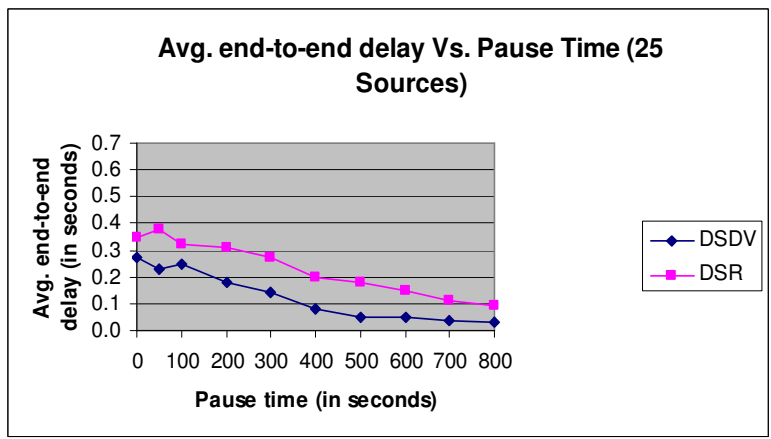

Figure 5. Average End to End Delay vs. Pause time for 25 Sources

From studying Figure4 and Figure5 for average end-to-end delay we see that DSDV has less delay in comparison to DSR. DSDV is a proactive routing protocol. In DSDV nodes periodically exchange routing tables between them in order to maintain up-to-date routing information to all destinations. Hence, whenever a source node wants to send a packet to a destination node, with the already available routing information it can do so without wasting any time for path setup. This reduces the average end-to-end delay of DSDV.

DSR on the other hand is a reactive source routing protocol. If a node in DSR wants to send a packet to a destination node, it has to find the route to the destination first. This route discovery latency is a part of the total delay. DSR being a source routing protocol, the initial path set up time is significantly higher. During the route discovery process every intermediate node needs to extract the information before forwarding the data packet. Moreover in DSR, the source has to wait for all the replies sent against every request reaching the destination. This increases the delay. While delivering a packet to a destination node, if DSR finds a link broken between two nodes on the path, it would make an attempt to find an alternate path from its cache entries, resulting in additional delay in packet delivery.

From the graphs we also see that the delay increases with increasing mobility and traffic as we increase the number of sources and reduce the pause time. As the mobility and traffic increase there will be more link breaks. The link breaks will necessitate new route discovery and thus increase the delay. Congestion will also be more with increasing mobility and traffic which also adds to the increasing delay.

Although DSR maintains multiple routes to the same destination in the cache, but it lacks any mechanism to determine the freshness of a route. It also does not have any mechanism to expire the stale routes. With high mobility link breaks become more frequent and there is the chance of the cached routes becoming stale quickly. DSR, being unable to determine a fresh route, may pick up a stale route for packet delivery. This unnecessarily consumes extra channel bandwidth and additional interface queue slots as the packet will ultimately be dropped. Moreover, every intermediate node can extract the information before forwarding the data packet and use this information to update its own cache entries. Therefore, selecting a stale route from a particular node's cache may pollute the cache entries of other nodes as well. This requires DSR to initiate more route discoveries which further adds to the increasing delay.

### 5.3. Normalized Routing Load Comparison

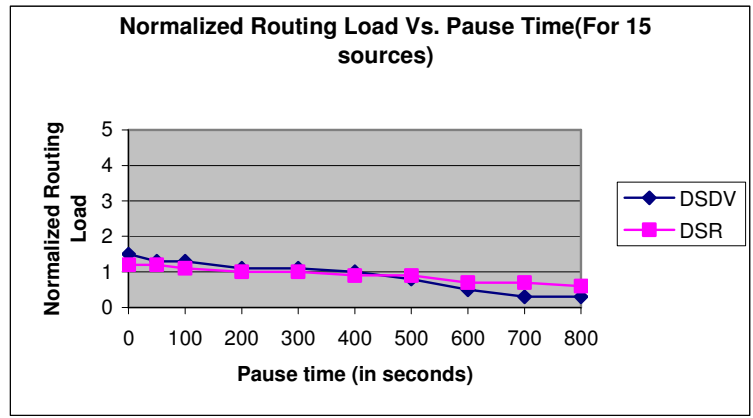

Figure 6. Normalized Routing Load vs. Pause Time for 15 Sources

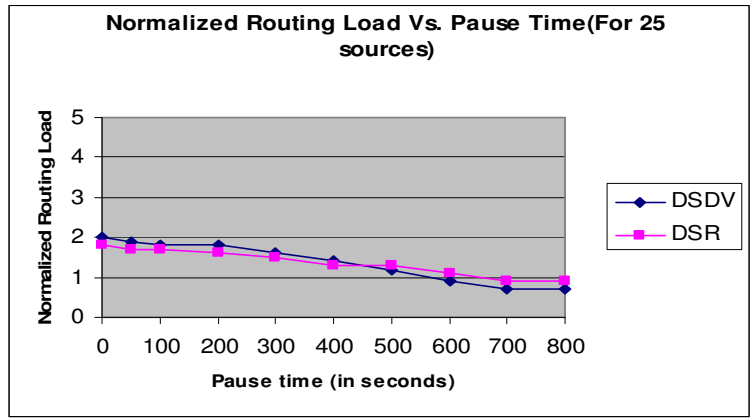

Figure 7. Normalized Routing Load vs. Pause Time for 25 Sources

From Figure6 and Figure7 we note that initially when the mobility is low, DSR has greater normalized routing load. This is attributed to the fact that DSR being a source routing protocol, with every packet the entire routing information is embedded. In addition to that, in response to a route discovery, replies come from many intermediate nodes. This increases the total control traffic. In case of DSDV, initially, when the mobility is low, the network topology remains relatively stable. Hence, nodes need to exchange only incremental dumps rather than full dumps. This results in lesser overhead of DSDV.

With higher mobility the network topology changes frequently. DSDV being proactive in nature, is less adaptive to this quickly changing scenario. Therefore, nodes need to exchange full dumps in order to maintain up-to-date routing information. This causes greater routing overhead for DSDV. In comparison, DSR uses aggressive caching strategy and the hit ratio is quite high. As a consequence, in high mobility scenario even if a link breaks, DSR can resort to an alternate link already available in the cache. Thus the route discovery process can be postponed until all the routes in the cache fail. This reduces the frequency of route discovery, which ultimately results in less routing overhead of DSR.

## 6. CONCLUSIONS

In this paper we have carried out a detailed ns2 based comparative simulation study of the performance characteristics of DSDV and DSR under hybrid scenario. Our work is the first in an attempt to compare these protocols in hybrid networking environment.

The simulation results show that at higher mobility DSR outperforms DSDV in terms of packet delivery performance. This is attributed to the DSR's ability to maintain multiple routes per destination and its use of aggressive caching strategy. At lower mobility, however, DSDV performs better than DSR. The network being relatively stable, at the time of packet delivery, all the routes are already available in DSDV due to its proactive nature. This results in greater packet delivery fraction.

Our experiment results also indicate that DSR exhibits more average end-to-end delay in comparison to DSDV. This is due to the fact that DSR being a source routing protocol, the initial path set up time is significantly higher as during the route discovery process every intermediate node needs to extract the information before forwarding the data packet. Although DSR maintains multiple routes to the same destination in the cache, but it lacks any mechanism to determine the freshness of the routes or to expire the stale routes. With high mobility and frequent link breaks there are chances of more routes becoming stale quickly. This requires the DSR to initiate the route discovery process which further adds to the increasing delay.

At higher mobility we note that DSR has lower routing load than DSDV. DSR uses aggressive caching technique and maintains multiple routes to the same destination. Hence, in high mobility scenario even if a link breaks, DSR can resort to an alternate link already available in the cache. This reduces the frequency of route discovery, which ultimately results in lower routing overhead of DSR. On the other hand, at lower mobility, the network topology remains relatively stable. Hence, in DSDV, nodes need to exchange only incremental dumps rather than full dumps. This results in lesser overhead of DSDV.

Thus we can conclude that if routing delay is of little concern, then DSR shows better performance at higher mobility in terms of packet delivery fraction and normalized routing load in hybrid networking scenario. Under less stressful scenario, however, DSDV outperforms DSR in terms of all three metrics.

While in this work we focus on the three prime metrics to analyze the performance of these protocols, there are many other issues that need to be considered to have an in-depth idea of these protocols' behaviour in hybrid networking environment. In our future work, we plan to study the performance of these protocols under other network scenarios by varying the network size, the number of connections, the mobility models and the speed of the mobile nodes etc.

**Authors**

**Koushik Majumder** has received his B.Tech and M.Tech degrees in Computer Science and Engineering and Information Technology in the year 2003 and 2005 respectively from University of Calcutta, Kolkata, India. Before coming to the teaching profession he has worked in reputed international software organizations like Tata Consultancy Services and Cognizant Technology Solutions. He is presently working as an Assistant Professor in the Dept. of Computer Science & Engineering in West Bengal University of Technology, Kolkata, India He is currently working towards his PhD degree in the Area of Mobile Ad hoc Networks. He has published several papers in International and National level journals and conferences.

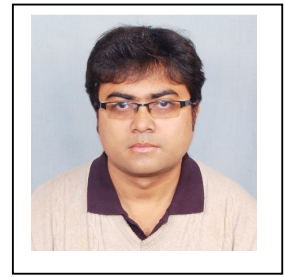

**Prof Subir Kumar Sarkar** received the B. Tech and M. Tech. Degree from the Institute of Radio Physics and Electronics, University of Calcutta in 1981 and 1983, respectively and PhD (Tech) degree in Microelectronics from University of Calcutta. He served Oil and Natural Gas Commission (ONGC) as an Executive Engineer for about 10 years( 1982 to 1992) before coming to teaching profession. He joined as a faculty member in the Dept. of Electronics and Telecommunication Engineering, Bengal Engineering and Science University, Shibpur in April 1992 (from 1992 to 1999). In 1999 he joined in Jadavpur University in the same dept. where he is presently a Professor. He has developed several short courses for the needs of the Engineers. He has published three Engineering text books and more than 250 technical research papers in archival journals and peer – reviewed conferences. His most recent research focus is in the areas of simulations of nanodevice models ,transport phenomenon, single electron & spintronics devices and their applications in VLSI circuits , ad hoc wireless networks, wireless mobile communication and data security. He is a Fellow of the Institution of Engineers, member of Indian Association for the cultivation of Science.

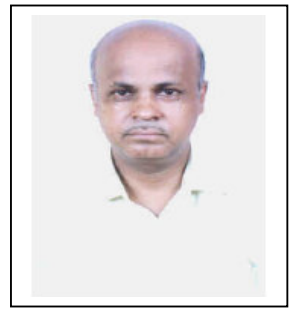